\documentclass[doublecol]{epl2}
\usepackage{amsmath,amssymb}
\usepackage{graphicx}

\title{Competing Ground States of a Peierls-Hubbard Nanotube}
\shorttitle{Competing Ground States of a Peierls-Hubbard Nanotube}
\author{Jun Ohara and Shoji Yamamoto}
\shortauthor{J. Ohara \etal}

\institute{Department of Physics, Hokkaido University,
         Sapporo 060-0810, Japan}

\abstract
{Motivated by iodo platinum complexes assembled within a
quadratic-prism lattice,
[Pt(C$_2$H$_8$N$_2$)(C$_{10}$H$_8$N$_2$)I]$_4$(NO$_3$)$_8$,
we investigate the ground-state properties of a Peierls-Hubbard
four-legged tube.
Making a group-theoretical analysis, we systematically reveal a variety
of valence arrangements, including half-metallic charge-density-wave
states.
Quantum and thermal phase competition is numerically demonstrated
with particular emphasis on doping-induced successive
insulator-to-metal transitions with conductivity increasing stepwise.}

\begin{document}

\maketitle

   Quasi-one-dimensional halogen ($X$)-bridged transition-metal ($M$)
complexes \cite{M5758,M5763,G6408,W6435} are unique optoelectronic
materials.
A platinum-chloride chain compound,
[Pt(ea)$_4$Cl]Cl$_2\cdot$2H$_2$O
(ea$\,=\,$ethylamine$\,=\,$C$_2$H$_7$N),
well-known as Wolffram's red salt, exhibits a Peierls-distorted
mixed-valent ground state \cite{C475}, whereas its nickel analog has a
Mott-insulating monovalent regular-chain structure \cite{T4261,T2341}.
Microscopic electronic-structure calculations demonstrated the robustness
\cite{R3913} and tunability \cite{A1415} of the Peierls instability.
Metal binucleation leads to a wider variety of electronic states
\cite{Y183,Y125124,K2163}.
Diplatinum-halide chain compounds,
$R_4$[Pt$_2$(pop)$_4X$]$\cdot n$H$_2$O
[$X=\,$Br, I;
 pop$\,=\,$diphosphonate$\,=\,$P$_2$O$_5$H$_2$;
 $R=\,$K, (C$_2$H$_5$)$_2$NH$_2$] \cite{C4604,K4420}, have a ground state
with halogen-sublattice dimerization, which is reminiscent of the $M\!X$
conventional, while their analog without any counter ion,
Pt$_2$(dta)$_4$I
(dta$\,=\,$dithioacetate$\,=\,$CH$_3$CS$_2$) \cite{B444,B2815}, possesses
a novel ground state with metal-sublattice dimerization, where twisting
of the dta ligand possibly plays an essential role \cite{B4562}.
The former exhibits photo- and/or pressure-induced phase transitions
\cite{S1405,Y140102,M046401,Y075113}, whereas the latter undergoes
successive phase transitions with increasing temperature
\cite{K10068,Y1198}.
There are further attempts \cite{S8366,Y6596} at bridging polynuclear
and/or heterometallic units by halogens.
\begin{figure}
\centering
\includegraphics[width=85mm]{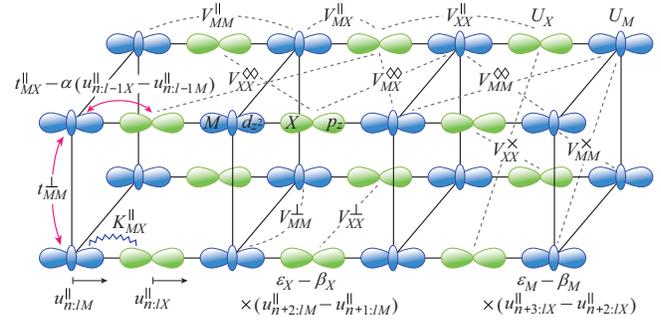}
\vspace*{-2mm}
\caption{(Color online)
         Modelling of an $M\!X$ quadratic prism, where heavily and
         lightly shaded clouds denote $M\,d_{z^2}$ and $X\,p_z$
         orbitals, the electron numbers on which are given by
         $n_{n:lMs}\equiv a_{n:lMs}^\dagger a_{n:lMs}$ and
         $n_{n:lXs}\equiv a_{n:lXs}^\dagger a_{n:lXs}$, respectively.
         The on-site energies of isolated atoms are given by
         $\varepsilon_M$ and $\varepsilon_X$,
         while the electron hoppings between these levels are modelled by
         $t_{M\!X}^{\parallel}$ and $t_{M\!M}^{\perp}$.
         The on-site Coulomb interactions are labelled as $U_A$
         ($A=M,X$), whereas the interchain and intrachain
         different-site Coulomb interactions as
         $V_{A\!A'}^{\perp}$,
         $V_{A\!A'}^{\lozenge\!\lozenge}$,
         $V_{A\!A'}^{\times}$, and
         $V_{A\!A'}^{\parallel}$ ($A,A'=M,X$).
         The leg-direction displacements of metal and halogen ions,
         $u_{n:lM}^{\parallel}$ and $u_{n:lX}^{\parallel}$, interact with
         electrons through intersite ($\alpha$) and intrasite
         ($\beta_M,\beta_X$) coupling constants at the cost of elastic
         energy $\propto K_{M\!X}^{\parallel}$.}
\label{F:H}
\end{figure}

   Hundreds of $M\!X$ compounds have thus been synthesized and studied,
but they all have single-chain-assembled structures.
In such circumstances, several chemists designed $M\!X$ ladders
\cite{K12066,K7372}.
Platinum-halide double-chain compounds,
($\mu$-bpym)[Pt(en)$X$]$_2X$(ClO$_4$)$_3\cdot$H$_2$O
($X=\,$Cl, Br;
 en$\,=\,$ethylendiamine$\,=\,$C$_2$H$_8$N$_2$;
 $\mu$-bpym$\,=2,2'$-bipyrimidine$\,=\,$C$_8$H$_6$N$_4$)
and
(bpy)[Pt(dien)Br]$_2$Br$_4\cdot 2$H$_2$O
(dien$\,=\,$diethylentriamine$\,=\,$C$_4$H$_{13}$N$_3$;
 bpy$\,=4,4'$-bipyridyl$\,=\,$C$_{10}$H$_8$N$_2$),
are made in distinct ground states of mixed valence \cite{F044717,I063708}
and they are optically distinguishable \cite{Y235116,Y367}.
Another chemical exploration is so exciting as to spark renewed
interest not only in $M\!X$ materials but also in the modern
microelectronics.
Otsubo and Kitagawa \cite{O} have patterned $M\!X$ chains in a nanotube
and fabricated a quadratic-prism compound,
[Pt(en)(bpy)I]$_4$(NO$_3$)$_8$.
Tubed metal complexes are scarcely precedented and serve as a new
laboratory distinct from $sp^2$-bonded-carbon nanotubes \cite{I56}.
The bpy ligands can be replaced in an attempt to tune the inside diameter,
while alternative bridging halide ions may enhance the Peierls distortion.
Platinum-halide tubes and ribbons potentially reveal fully correlated
electrons coupled with phonons on a way from one to two dimensions.
The cylindrical structure may yield novel valence arrangements of its own
and unlikely in an open plain.
A theoretical scenario for quantum, thermal, and possibly photoinduced
transitions between them must stimulate extensive experimental
explorations of this new $M\!X$ family.

   Thus motivated, we investigate broken-symmetry solutions of a
four-legged Peierls-Hubbard tube.
A group-theoretical bifurcation theory predicts the variety of ground
states in a platinum-halide quadratic prism.
Numerical calculations visualize their close competition as a function of
temperature, electron occupancy, and Coulomb interactions.
Determination of any one-dimensional structure demands an elaborate
analysis \cite{W6676} of the diffuse X-ray scattering intensity.
Resonant Raman spectroscopy \cite{K4420} is potentially eloquent of the
valence arrangement.
Our systematic analysis based on a symmetry argument will stimulate and
serve for such structural investigations.

   Metal-halide quadratic prisms are describable with a two-band
extended Peierls-Hubbard Hamiltonian,
\begin{eqnarray}
   &&\!\!\!\!\!\!\!\!
   {\cal H}=
   \sum_{l,n,s}
   \Bigl\{
    \bigl[t_{M\!X}^{\parallel}
         -\alpha(u_{n+1:lM}^{\parallel}-u_{n:lX}^{\parallel})\bigr]
    a^{\dagger}_{n+1:lMs}a_{n:lXs}
   \nonumber \\
   &&\!\!\!\!\!\!\!\!\qquad\quad
   -\bigl[t_{M\!X}^{\parallel}
         -\alpha(u_{n:lX}^{\parallel}-u_{n:lM}^{\parallel})\bigr]
    a^{\dagger}_{n:lXs}a_{n:lMs}
   \nonumber \\
   &&\!\!\!\!\!\!\!\!\qquad\quad
   -t_{M\!M}^{\perp}
    a^{\dagger}_{n:l+1Ms}a_{n:lMs}
   +{\rm H.c.}
   \Bigr\}
   \nonumber \\
   &&\!\!\!\!\!\!\!\!\quad
  +\sum_{l,n,s}
   \Bigl\{
    \bigl[\varepsilon_{M}
         -\beta_{M}(u_{n:lX}^{\parallel}-u_{n-1:lX}^\parallel)\bigr]
    n_{n:lMs}
   \nonumber \\
   &&\!\!\!\!\!\!\!\!\qquad\quad
   +\bigl[\varepsilon_{X}
         -\beta_{X}(u_{n+1:lM}^{\parallel}-u_{n:lM}^{\parallel})\bigr]
    n_{n:lXs}
   \Bigr\}
   \nonumber \\
   &&\!\!\!\!\!\!\!\!\quad
  +\sum_{l,n}\frac{K_{M\!X}^{\parallel}}{2}
    \bigl[(u_{n:lX}^{\parallel}-u_{n:lM}^{\parallel})^{2}
         +(u_{n+1:lM}^{\parallel}-u_{n:lX}^{\parallel})^{2}\bigr]
   \nonumber \\
   &&\!\!\!\!\!\!\!\!\quad
  +\sum_{A=M,X}\sum_{l,n,s,s'}
   \Bigl\{
    \frac{U_{A}}{4}n_{n:lAs}n_{n:lA-s}
   +V_{A\!A}^{\parallel}n_{n:lAs}n_{n+1:lAs'}
   \nonumber \\
   &&\!\!\!\!\!\!\!\!\qquad\quad
   +V_{A\!A}^{\perp}n_{n:lAs}n_{n:l+1As'}
   +\frac{V_{A\!A}^{\times}}{2}n_{n:lAs}n_{n:l+2As'}
   \nonumber \\
   &&\!\!\!\!\!\!\!\!\qquad\quad
   +V_{A\!A}^{\lozenge\!\lozenge}
    (n_{n:lAs}n_{n+1:l+1As'}+n_{n:l+1As}n_{n+1:lAs'})
   \Bigr\}
   \nonumber \\
   &&\!\!\!\!\!\!\!\!\quad
  +\sum_{l,n,s,s'}
   \Bigl\{
    V_{M\!X}^{\parallel}
    (n_{n:lMs}n_{n:lXs'}+n_{n:lXs}n_{n+1:lMs'})
   \nonumber \\
   &&\!\!\!\!\!\!\!\!\quad
   +V_{M\!X}^{\lozenge\!\lozenge}
    (n_{n:lMs}n_{n:l+1Xs'}+n_{n:l+1Ms}n_{n:lXs'}
   \nonumber \\
   &&\!\!\!\!\!\!\!\!\qquad\quad
    +n_{n:lXs}n_{n+1:l+1Ms'}+n_{n:l+1Xs}n_{n+1:lMs'})
   \Bigr\},
   \label{E:H}
\end{eqnarray}
as is illustrated with Fig. \ref{F:H}, where $M\!X$ chain legs and
$M_4X_4$ units of rectangular parallelepiped are numbered by
$l=1,\cdots,4$ and $n=1,\cdots,N$, respectively, while electron spins
are indicated by $s,s'=\uparrow,\downarrow$.
\begin{table}
\caption{Axial isotropy subgroups and their fixed-point subspaces for
         the irreducible representations
         ${\rm X}\check{D}({\rm X})
          \otimes\check{S}^0\otimes\check{T}^0$.}
\begin{tabular}{cccc}
\hline \hline
$\check{D}({\rm X})$
& \hspace{-2mm}
Axial isotropy subgroup
& \hspace{-3mm}
$\begin{matrix}
\text{Fixed-point} \\[-1mm]
\text{subspace}
\end{matrix}$
\\
\hline\\[-4mm]
  $A_{1g}$
& \hspace{-2mm}$\mathbf{D}_{\rm 4h}\mathbf{L}_{2}\mathbf{ST}$
& \hspace{-3mm}$h_{XA_{1g}[1,1]}^{00}$ \\
  $A_{2g}$
& \hspace{-2mm}$(1+C_{2x}l)\mathbf{C}_{\rm 4h}\mathbf{L}_{2}\mathbf{ST}$
& \hspace{-3mm}$h_{XA_{2g}[1,1]}^{00}$ \\
  $B_{1g}$
& \hspace{-2mm}$(1+C_{2a}l)\mathbf{D}_{2h}\mathbf{L}_{2}\mathbf{ST}$
& \hspace{-3mm}$h_{XB_{1g}[1,1]}^{00}$ \\
  $B_{2g}$
& \hspace{-2mm}$(1+C_{2x}l)\mathbf{D}_{2ah}\mathbf{L}_{2}\mathbf{ST}$
& \hspace{-3mm}$h_{XB_{2g}[1,1]}^{00}$ \\
  $E_{g}^{(1)}$
& \hspace{-2mm}$(1+C_{2z}l)\mathbf{C}_{2xh}\mathbf{L}_{2}\mathbf{ST}$
& \hspace{-3mm}$h_{XE_{g}[1,1]}^{00}$ \\
  $E_{g}^{(2)}$
& \hspace{-2mm}$(1+C_{2z}l)\mathbf{C}_{2ah}\mathbf{L}_{2}\mathbf{ST}$
& \hspace{-3mm}$\sum_{i,j=1}^{2}h_{XE_{g}[i,j]}^{00}/2$ \\
  $A_{1u}$
& \hspace{-2mm}$(1+Il)\mathbf{D}_{4}\mathbf{L}_{2}\mathbf{ST}$ 
& \hspace{-3mm}$h_{XA_{1u}[1,1]}^{00}$ \\
  $A_{2u}$
& \hspace{-2mm}$(1+Il)(1+IC_{2x})\mathbf{C}_{4}\mathbf{L}_{2}\mathbf{ST}$
& \hspace{-3mm}$h_{XA_{2u}[1,1]}^{00}$ \\
  $B_{1u}$
& \hspace{-2mm}$(1+Il)(1+IC_{2a})\mathbf{D}_{2}\mathbf{L}_{2}\mathbf{ST}$ 
& \hspace{-3mm}$h_{XB_{1u}[1,1]}^{00}$ \\
  $B_{2u}$
& \hspace{-2mm}$(1+Il)(1+IC_{2x})\mathbf{D}_{2a}\mathbf{L}_{2}\mathbf{ST}$
& \hspace{-3mm}$h_{XB_{2u}[1,1]}^{00}$ \\
  $E_{u}^{(1)}$
& \hspace{-2mm}$(1+Il)(1+IC_{2y})\mathbf{C}_{2x}\mathbf{L}_{2}\mathbf{ST}$
& \hspace{-3mm}$h_{XE_{u}[1,1]}^{00}$ \\
  $E_{u}^{(2)}$
& \hspace{-2mm}$(1+Il)(1+IC_{2b})\mathbf{C}_{2a}\mathbf{L}_{2}\mathbf{ST}$
& \hspace{-3mm}$\sum_{i,j=1}^{2}h_{XE_{u}[i,j]}^{00}/2$ \\[1mm]
\hline \hline
\end{tabular}
\label{T:Rrep}
\end{table}
\begin{figure*}
\centering
\includegraphics[width=170mm]{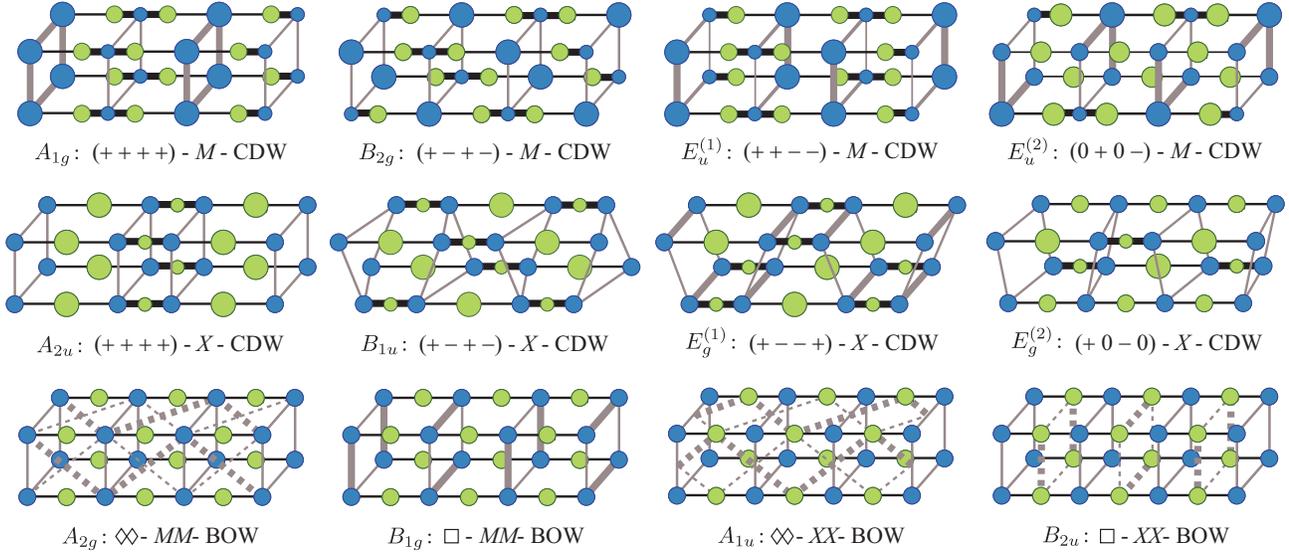}
\vspace*{-2mm}
\caption{(Color online)
         Possible density waves of the
         ${\rm X}\check{D}({\rm X})\otimes\check{S}^0\otimes\check{T}^0$
         type, where varied circles and segments represent oscillating
         electron densities and bond orders, respectively, while
         irregularly arranged circles denote lattice distortion.
         Various $M$ ($X$)-CDW states are referred to as
         $(\sigma_1,\sigma_2,\sigma_3,\sigma_4)$, where the signatures
         $\sigma_l=\pm,0$ denote the charges relative to $M^{3+}$ ($X^-$)
         on adjacent metal (halogen) sites forming a square section of the
         quadratic prism.}
\label{F:DW}
\end{figure*}

   When we consider normal states, the symmetry group of any lattice
electron system may be written as
$\mathbf{G}=\mathbf{P}\times\mathbf{S}\times\mathbf{T}$,
where $\mathbf{P}$, $\mathbf{S}$, and $\mathbf{T}$ are the groups of
space, spin rotation, and time reversal, respectively.
The space group is further decomposed into the translation and point
groups as $\mathbf{L}\land\mathbf{D}$.
For the present $d_{z^2}$-$p_z$ quadratic prism, $\mathbf{L}$ and
$\mathbf{D}$ read as $\{E,l\}\equiv\mathbf{L}_1$ and $\mathbf{D}_{\rm 4h}$,
respectively, where $l$ is the unit-cell translation in the $z$
direction.
Defining the Fourier transformation as
$a_{k:lAs}
 =N^{-1/2}\sum_n e^{-ik(n+\delta_{AX}/2)}a_{n:lAs}$ and
$u_{k:lA}^{\parallel}
 =N^{-1/2}\sum_n e^{-ik(n+\delta_{AX}/2)}u_{n:lA}^{\parallel}$ with the
lattice constant along the legs set equal to unity and composing
Hermitian bases of the gauge-invariant operators
$\{a_{k:lAs}^{\dagger}a_{k':l'A's'}\}$,
we investigate irreducible representations of $\mathbf{G}$ over the
real number field, which are referred to as $\check{G}$.
Actions of $l\in\mathbf{L}_1$ and $t\in\mathbf{T}$ on the electron
operators are defined as
$l\cdot a_{k:lAs}^\dagger=e^{-ikl}a_{k:lAs}^\dagger$ and
$t\cdot a_{k:lAs}^\dagger
 =(-1)^{\delta_{s\uparrow}}a_{-k:lA-s}^\dagger$.
Those of $p\in\mathbf{D}_{\rm 4h}$ are calculated as
$p\cdot a_{k:lMs}^\dagger=[A_{1g}(p)]_{11}a_{pk:lMs}^\dagger$ and
$p\cdot a_{k:lXs}^\dagger=[A_{2u}(p)]_{11}a_{pk:lXs}^\dagger$, where
$[\check{D}(p)]_{ij}$ is the $(i,j)$-element of the $\check{D}$
representation matrix for $p$.
Those of
$u(\mbox{\boldmath$e$},\theta)
 =\sigma^0\cos(\theta/2)
 -i(\mbox{\boldmath$\sigma$}\cdot\mbox{\boldmath$e$})
  \sin(\theta/2)
 \in\mathbf{S}$
read as
$u(\mbox{\boldmath$e$},\theta)\cdot a_{k:lAs}^\dagger
 =\sum_{s'}[u(\mbox{\boldmath$e$},\theta)]_{s's}a_{k:lAs'}^\dagger$,
where $\sigma^0$ and
$\mbox{\boldmath$\sigma$}=(\sigma^x,\sigma^y,\sigma^z)$
are the $2\times 2$ unit matrix and a vector composed of the Pauli
matrices, respectively.
Any representation $\check{G}$ is expressed as
$\check{G}=\check{P}\otimes\check{S}\otimes\check{T}$.
Once a wave vector $Q$ is fixed, the relevant little group
$\mathbf{D}(Q)$ is given.
$\check{P}$ is therefore labelled as $Q\check{D}(Q)$.
The relevant representations of $\mathbf{S}$ are given by
$\check{S}^0(u(\mbox{\boldmath$e$},\theta))
 =1$ (singlet) and
$\check{S}^1(u(\mbox{\boldmath$e$},\theta))
 =O(u(\mbox{\boldmath$e$},\theta))$ (triplet),
where $O(u(\mbox{\boldmath$e$},\theta))$ is the $3\times 3$
orthogonal matrix satisfying
$
   u(\mbox{\boldmath$e$},\theta)
   \mbox{\boldmath$\sigma$}^\lambda
   u^\dagger(\mbox{\boldmath$e$},\theta)
      =\sum_{\mu=x,y,z}
       [O(u(\mbox{\boldmath$e$},\theta))]_{\lambda\mu}
       \mbox{\boldmath$\sigma$}^\mu \ \
       (\lambda=x,\,y,\,z)
$,
while those of $\mathbf{T}$ by
$\check{T}^0(t)=1$ (symmetric) and $\check{T}^1(t)=-1$ (antisymmetric).
Halogen-bridged platinum complexes are describable with moderately
correlated electrons, where neither magnetically ordered phase nor
current-wave state has ever been observed without any external field
applied.
The relevant $d$ and $p$ bands of an as-grown platinum-iodide
quadratic-prism compound are of $3/4$ electron filling.
Thus and thus, we discuss nonmagnetic solutions labelled
$\Gamma\check{D}(\Gamma)\otimes\check{S}^0\otimes\check{T}^0$ and
${\rm X}\check{D}({\rm X})\otimes\check{S}^0\otimes\check{T}^0$, where
$\Gamma$ and X denote $Q=0$ and $Q=\pi$, respectively, and their space
groups read as $\mathbf{L}_1\land\mathbf{D}_{\rm 4h}$ \cite{Y949} and
$\mathbf{L}_2\land\mathbf{D}_{\rm 4h}$ \cite{O55}, respectively, with
$\mathbf{L}_1\equiv\{E,l\}$ and $\mathbf{L}_2\equiv\{E,2l\}$.

   Then the Hamiltonian (\ref{E:H}) may be rewritten within the
Hartree-Fock scheme as
\begin{eqnarray}
   &&\!\!\!\!
   {\cal H}_{\rm HF}
   =\sum_{l,l'}\sum_{A,A'}
    \sum_{Q=\Gamma,{\rm X}}\sum_{k,s,s'}
    \sum_{\lambda=0,x,y,z}
    x_{lAl'A'}^\lambda(Q;k)
   \nonumber\\
   &&\qquad\times
    a_{k+Q:lAs}^\dagger a_{k:l'A's'}\sigma_{ss'}^\lambda
   \equiv\sum_{Q}\sum_\lambda h_Q^\lambda,
   \label{E:HHF}
\end{eqnarray}
where the order parameters $x_{lAl'A'}^\lambda(Q;k)$, as well as
the lattice distortion $u_{Q:lA}^{\parallel}$, should be determined so
as to minimize the free energy at every temperature given.
Employing the projection operators
\begin{equation}
   P_{\check{D}[i,j]}^\tau
  =\frac{d_{\check{D}}}{2g}
    \sum_{t\in\mathbf{T}}\check{T}^\tau(t)
    \sum_{p\in\mathbf{D}_{\rm 4h}}[\check{D}(p)]_{ij}^*tp,
\end{equation}
where
$g\,(=16)$ is the order of $\mathbf{D}_{\rm 4h}$ and
$d_{\check{D}}\,(\leq 2)$ is the dimension of its arbitrary irreducible
representation $\check{D}$, we further decompose the Hamiltonian
(\ref{E:HHF}) into symmetry-definite irreducible components
\cite{Y949,O55} as
\begin{equation}
   \!\!\!\!
   {\cal H}_{\rm HF}
   =\sum_{Q=\Gamma,{\rm X}}
    \sum_{\check{D}(Q)}
    \sum_{\lambda=0,x,y,z}
    \sum_{\tau=0,1}
    h_{Q\check{D}(Q)}^{\lambda\tau}.
\end{equation}
We list in Table \ref{T:Rrep} the irreducible representations
${\rm X}\check{D}({\rm X})\otimes\check{S}^0\otimes\check{T}^0$ whose
isotropy subgroups are axial, together with their fixed-point subspaces
$h_{{\rm X}\check{D}({\rm X})}^{\lambda\tau}$, where
$h_{{\rm X}\check{D}[i,j]}^{\lambda\tau}
=P_{\check{D}[i,j]}^\tau\cdot h_{\rm X}^\lambda$.
All the one-dimensional isotropy subgroups are proved to give stable
solutions \cite{G83}.
Considering that the density matrices
$\rho_{l'A'lA}^\lambda(Q;k)
 =\sum_{s,s'}\langle a_{k+Q:lAs}^\dagger a_{k:l'A's'}\rangle_T
  \sigma_{ss'}^\lambda/2$, where
$\langle\cdots\rangle_T$ denotes the thermal average in a Hartree-Fock
eigenstate, are of the same symmetry as their host Hamiltonian, we
learn the oscillating pattern of charge densities
$\sum_s\langle a_{n:lAs}^\dagger a_{n:lAs}\rangle_T$ and bond orders
$\mbox{Re}\sum_s\langle a_{n:lAs}^\dagger a_{n':l'A's}\rangle_T$.
The consequent density-wave solutions of $Q={\rm X}$ are shown in
Fig. \ref{F:DW}.
While we have analyzed and calculated those of $Q=\Gamma$ as
well, none of them but the paramagnetic metal of the full symmetry
$\mathbf{D}_{\rm 4h}\mathbf{L}_{1}\mathbf{ST}$,
labelled as $\Gamma A_{1g}\otimes\check{S}^0\otimes\check{T}^0$
and referred to as PM, plays the ground state under any realistic
parametrization.

   The ${\rm X}\check{D}({\rm X})\otimes\check{S}^0\otimes\check{T}^0$
solutions are classified into three groups:
charge density waves on the metal sublattice
with the halogen sublattice distorted,
charge density waves on the halogen sublattice
with the metal sublattice distorted, and
bond order waves without any charge oscillation,
which are abbreviated as $M$-CDW, $X$-CDW, and BOW, respectively.
No lattice distortion accompanies BOW within the present Hamiltonian
(\ref{E:H}).
Every BOW state may be stabilized by direct electron transfers on the
oscillating bonds and their interactions with phonons, but any is of
little occurrence under realistic modelling.
There are twice four kinds of CDW states.
Although all the CDW states gain a condensation energy due to their
Peierls distortion, they are not necessarily gapped.
$M$- and $X$-CDW of the $(0+0-)$ type are {\it half metallic},
where two legs are valence-delocalized, while the rest are
valence-trapped.
Such states as cell-doubled but partially metallic are generally
possible in tubed $M\!X$ compounds, including triangular prisms,
whose little groups $\mathbf{D}({\rm X})$ all have a two-dimensional
irreducible representation of axial isotropy subgroup.
All the other CDW states are {\it fully gapped} at the boundaries of the
reduced Brillouin zone.
Since the $M\,d_{z^2}$ orbitals are half filled and lie higher in energy
than the fully occupied $X\,p_z$ orbitals, $\pi$-modulated $d$-electron
CDW states are most likely to appear in undoped samples at low
temperatures.

   Now we are eager to observe actual phase competitions.
We have many unknown electronic correlation parameters as well as
well-established crystallographic ones \cite{O}.
Then, extending the Ohno relationship \cite{O219} to our heteroatomic
system \cite{C341}, we design, unless otherwise noted, the Coulomb
interaction between different sites $n:lA$ and $n':l'A'$ as
$\bar{U}
/\kappa\sqrt{1+[4\pi\epsilon_0\bar{U}r_{n:lA;n':l'A'}/e^2]^2}$,
where
$\bar{U}$ is the averaged on-site Coulomb repulsion $(U_M+U_X)/2$,
$r_{n:lA;n':l'A'}$ the distance between the two sites
under no deformation,
$e$ the electron charge,
$\epsilon_0$ the vacuum dielectric constant, and
$\kappa$ the relative permittivity.
Considering x-ray diffraction measurements on the quadratic-prism
compound [Pt(en)(bpy)I]$_4$(NO$_3$)$_8$ \cite{O}, we stand on
$r_{n:lM;n+1:lM}=2r_{n:lM;n:lX}=6\,\mbox{\AA}$ and
$r_{n:lM;n:l+1M}=r_{n:lX;n:l+1X}=11\,\mbox{\AA}$, whereas
referring to optical investigations on the analogous ladder compound
(bpy)[Pt(dien)Br]$_2$Br$_4\cdot 2$H$_2$O \cite{Y235116,Y367},
we assume that
$t_{M\!X}^\parallel=1.5\,\mbox{eV}$, $t_{M\!M}^\perp=0.32\,\mbox{eV}$,
$U_M=1.2\,\mbox{eV}$, $U_X=1.0\,\mbox{eV}$,
$\varepsilon_M-\varepsilon_X=1.2\,\mbox{eV}$,
$\alpha=0.84\,\mbox{eV/\AA}$,
$\beta_M=\beta_X=2.3\,\mbox{eV}/\mbox{\AA}$, and
$K_{M\!X}^\parallel=8.0\,\mbox{eV}/\mbox{\AA}^2$.
Such a parametrization is consistent with previous model studies
\cite{G6408,W6435,C823}, first-principle calculations \cite{A2739}, and
photostructural investigations \cite{M046401,K1789,D3285} on $M\!X$ and
$M\!M\!X$ chains.
Under little information about platinum-halide dielectric constants,
we set $\kappa$ two ways, that is, equal to $2$, considering the moderate
screening in organic semiconductors \cite{Y266222,Y235205}, and equal to
$4$, considering the strong screening in transition-metal complexes
\cite{C3985}.
\begin{figure}
\centering
\includegraphics[width=85mm]{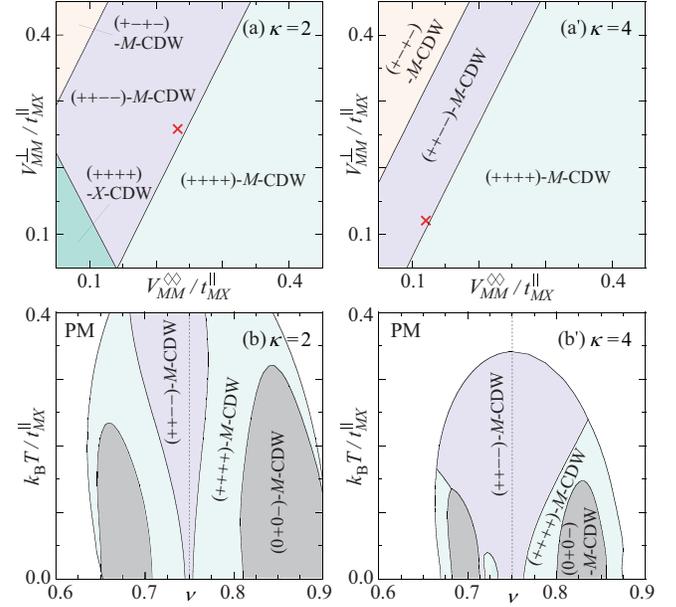}
\vspace*{-2mm}
\caption{(Color online)
         Ground-state phase diagrams on the
         $V_{M\!M}^{\perp}$-$V_{M\!M}^{\lozenge\!\lozenge}$ square,
         where the crosses indicate the Ohno-type parametrizations with
         $\kappa=2$ (a) and $\kappa=4$ (${\rm a}'$) and we move away from
         these points tuning only $V_{M\!M}^{\perp}$ and
         $V_{M\!M}^{\lozenge\!\lozenge}$, and
         thermal phase diagrams with varying electron occupancy $\nu$
         under the $\kappa=2$ (b) and $\kappa=4$ (${\rm b}'$) Ohno-type
         parametrizations, where the dotted lines are guides for eyes,
         separating the hole- and electron-doped regions.}
\label{F:PhD}
\end{figure}

   Figure \ref{F:PhD}(a) demonstrates quantum phase transitions in the
low-temperature limit.
The competition within $M$-CDW states is straightforwardly
understandable when we assume the $X\,p_z$ orbitals to be fully filled
and thus inactive.
Those of $(++++)$, $(+-+-)$ and $(++--)$ are stabilized
with increasing $V_{M\!M}^{\lozenge\!\lozenge}$, $V_{M\!M}^{\perp}$,
and $V_{M\!M}^{\times}$, respectively.
Interchain electron transfers also bring about energy gains in all but
the first.
The phase boundaries are roughly given by
$V_{M\!M}^{\perp}=2V_{M\!M}^{\lozenge\!\lozenge}\pm V_{M\!M}^{\times}$
under slight correction $\propto(t_{M\!M}^{\perp})^2$.
$X$-CDW states are of occurrence with $p$ electrons strongly
correlating against the relative electron affinity
$\varepsilon_X-\varepsilon_M$.
The present parametrizations both suggest an $M$-CDW ground state of the
$(++--)$ type closely competing with that of the $(++++)$ type.
From the theoretical point of view, $(++--)$-$M$-CDW is
characteristic of a tubed lattice in that it belongs to a
two-dimensional representation.
On the other hand, $(++++)$-$M$-CDW and $(+-+-)$-$M$-CDW have
good analogy with CDW states of the in-phase (IP) and out-of-phase (OP)
types, respectively, found in $M\!X$ ladder compounds \cite{Y235116}.
We are all excited at the thought of structural investigations of
[Pt(en)(bpy)I]$_4$(NO$_3$)$_8$.

\begin{figure}
\centering
\includegraphics[width=85mm]{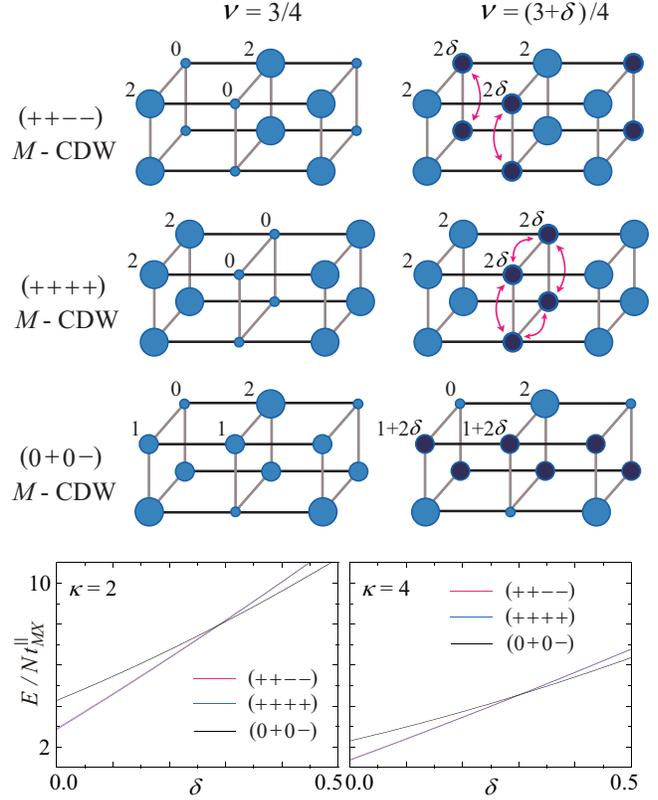}
\vspace*{-2mm}
\caption{(Color online)
         Electron occupancy of the ${\rm Pt}\,d_{z^2}$ orbitals under
         electron doping in the strong-intrasite-coupling limit.
         Doped electrons are evenly distributed among the four chains in
         $(++++)$-$M$-CDW and $(++--)$-$M$-CDW, whereas they are
         predominantly put into the two valence-delocalized paramagnetic
         chains in $(0+0-)$-$M$-CDW.
         The bent arrows signify doping-induced energy gains
         $\propto t_{M\!M}^{\perp}$.
         Energy estimates (\ref{E:++--})-(\ref{E:0+0-}) are plotted as
         functions of $\delta\geq 0$ at $\kappa=2$ and $\kappa=4$.}
\label{F:doping}
\end{figure}

   Figure \ref{F:PhD}(b) stimulates another interest in platinum-halide
prism compounds.
The $(++--)$-to-$(++++)$ transition with the electron occupancy
$\nu$ moving away from $3/4$ is caused by activated interchain
electron hopping.
Under the present Coulomb parametrizations, 
$(++--)$-$M$-CDW and $(++++)$-$M$-CDW are almost balanced at
$\nu=3/4$ and it is the slight energy correction
$\propto(t_{M\!M}^{\perp})^2$ that stabilizes the former over the
latter.
There is no interchain electronic communication between phased
$3/4$-filled CDW chains in the strong-coupling limit.
However, slightly doped electrons or holes bring about energy gains
$\propto t_{M\!M}^{\perp}$ in $(++++)$-$M$-CDW as well as in
$(++--)$-$M$-CDW, which are illustrated with bent arrows in Fig.
\ref{F:doping}, and those in the former are roughly twice as much as
those in the latter.
That is why $(++++)$-$M$-CDW is quick to replace
$(++--)$-$M$-CDW under doping.
Further doping destabilizes the $\pi$-modulated Peierls distortion
and induces a quite interesting phase, $(0+0-)$-$M$-CDW, which is
derived from another two-dimensional representation,
${\rm X}E_{u}^{(2)}\otimes\check{S}^0\otimes\check{T}^0$.
There occurs a partially metallic state in between the totally
valence-trapped and fully metallic states, where intermediate
conductivity should be observed.
A thermal transition to PM is hardly realistic considering
$t_{M\!X}^\parallel$ of eV order, while doping-induced quantum
transitions to metallic states may be feasible.
Electrochemical doping, by exposing a single crystal to halogen vapor
\cite{H5706}, for instance, possibly causes successive phase transitions
towards the fully metallic state with conductivity increasing stepwise.

   The doping-induced stabilization of the novel half-metallic phase
against the fully distorted $M$-CDW states is well understandable within a
simple consideration of their electronic correlation energies.
Figure \ref{F:doping} gives a single-band description of $M$-CDW states
under doping.
In  $(++--)$-$M$-CDW and $(++++)$-$M$-CDW, electrons are doped into vacant
sites, whereas holes into fully occupied sites.
In  $(0+0-)$-$M$-CDW, the metallic chains are doped first, while the
Peierls-distorted chains remain half-filled (in the single-band picture),
because the metallic Pt$^{3+}$ bands are sandwiched between the bonding
Pt$^{2+}$ and antibonding Pt$^{4+}$ bands far apart from them.
Their per-unit $d$-electron energies under electron doping are estimated
as
\begin{eqnarray}
   &&\!\!\!\!\!\!\!\!\!\!
   \frac{E^{(++--)}}{N}=
   2(1+\delta^2)U_{M}
  +8\delta(2V_{M\!M}^{\parallel}+V_{M\!M}^{\times})
   \nonumber \\
   &&\!\!\!\!\!\!\!\!\!\!\qquad\quad
  +4(1+\delta)^2(V_{M\!M}^{\perp}+2V_{M\!M}^{\lozenge\!\lozenge})
  -\frac{4\beta_{M}^{2}}{K_{M\!X}^{\parallel}}(1-\delta)^2,
   \label{E:++--} \\
   &&\!\!\!\!\!\!\!\!\!\!
   \frac{E^{(++++)}}{N}=
   2(1+\delta^2)U_{M}
  +16\delta(V_{M\!M}^{\parallel}+2V_{M\!M}^{\lozenge\!\lozenge})
   \nonumber \\
   &&\!\!\!\!\!\!\!\!\!\!\qquad\quad
  +4(1+\delta^2)(V_{M\!M}^{\times}+2V_{M\!M}^{\perp})
  -\frac{4\beta_{M}^{2}}{K_{M\!X}^{\parallel}}(1-\delta)^2,
   \label{E:++++} \\
   &&\!\!\!\!\!\!\!\!\!\!
   \frac{E^{(0+0-)}}{N}=
   (\frac{3}{2}+2\delta+2\delta^2)U_{M}
  +(1+2\delta)^2(2V_{M\!M}^{\parallel}+V_{M\!M}^{\times})
   \nonumber \\
   &&\!\!\!\!\!\!\!\!\!\!\qquad\quad
  +4(1+2\delta)(V_{M\!M}^{\perp}+2V_{M\!M}^{\lozenge\!\lozenge})
  -\frac{2\beta_{M}^{2}}{K_{M\!X}^{\parallel}},
   \label{E:0+0-}
\end{eqnarray}
and are visualized in Fig. \ref{F:doping}.
$(++--)$-$M$-CDW and $(++++)$-$M$-CDW are indeed closely competing with
each other and the most interesting $(++++)$-to-$(0+0-)$ transition is
reproduced well.
$E^{(++++)}$ and $E^{(0+0-)}$ are balanced at $\delta\simeq 0.29$ and
$\delta\simeq 0.30$ in the cases of $\kappa=2$ and $\kappa=4$,
respectively.
Because of the electron-hole symmetry in the single-band modelling, the
transition points under hole doping are simply obtained as $-\delta$.

   With respect to the appearance of a partially distorted prism lattice,
we should further note that any structural instability is
{\it conditional} in the present system.
Indeed a single $M\!X$ chain is unconditionally distorted
\cite{B339,Y165113}, but coupled $M\!X$ chains, whether tubed or not,
are never distorted under infinitesimal coupling.
It is the case with organic polymers as well.
The Peierls instability in polyacetylene is unconditional, whereas those
in polyacene are conditional \cite{Y235205,S435}.
It is not only due to Coulomb correlations but also of geometric origin
that half the chains remain undistorted in $(0+0-)$-$M$-CDW.

   Structural instabilities of longer period may also be mentioned in this
context.
We have indeed found CDW solutions of $0<Q<\pi$ under doping.
At $\nu=3/4\pm 1/16$, for example, there exists a quadratic prism composed
of two $3/4$-filled dimerized and two $(3/4\pm 1/8)$-filled tetramerized
chains as well as a wholly octamerized prism.
However, they are generally higher in energy than $(0+0-)$-$M$-CDW under
the present parametrizations.
Besides $2k_{\rm F}$ instabilities, $4k_{\rm F}$-CDW states such as all
the chains octamerized at $\nu=3/4\pm 1/32$ have also been found, but they
are inferior to $(++++)$-$M$-CDW in energy.
All such instabilities are conditional and the critical coupling strength
is on the whole an increasing function of the number of constituent chains
and the spatial period of oscillation.
There is a possibility \cite{B13228} of long-period ground states
appearing with stronger on-site electron-phonon coupling and/or
weaker intersite Coulomb interaction.
However, it may not be the case with our platinum-halide prisms,
especially with iodo complexes.
The Peierls gap
$\propto\beta_{M}/\sqrt{t_{M\!X}^{\parallel}K_{M\!X}^{\parallel}}$
decreases in the order ${\rm Cl}>{\rm Br}>{\rm I}$ \cite{T2212}, while
the IP-CDW ground states \cite{Y235116} of similar ladder compounds
($\mu$-bpym)[Pt(en)$X$]$_2X$(ClO$_4$)$_3\cdot$H$_2$O
demonstrate the relevance of the intersite Coulomb interactions.
We are hoping for large-scale measurement and further tuning of tubed
$M\!X$ compounds.

\acknowledgments

   We express special thanks to M. Ozaki for helpful comments on our
calculation and to K. Otubo and H. Kitagawa for valuable information on
their brandnew quadratic-prism $M\!X$ complexes.
This work was supported by the Ministry of Education, Culture, Sports,
Science, and Technology of Japan.

\end{document}